\documentclass [11pt,a4paper]{article}

\usepackage{shrthnds}
\usepackage{cite}
\usepackage{graphicx}
\usepackage{color}
\usepackage{subfigure}

\usepackage{multirow}

\usepackage[hang,small]{caption}

\usepackage{geometry}
\usepackage{titlesec,titletoc}
  \titleformat{\section}{\Large\bf}{\thesection.}{1em}{}
  \titleformat{\subsection}{\large\bf}{\thesubsection.}{1em}{}

\title{\bf One Loop Calculation of Cosmological Constant in a Scale Invariant Theory }

\author{\bf Pankaj Jain$^{1}$\footnote{email: pkjain@iitk.ac.in}~and~Subhadip Mitra$^{2}$\footnote{email: subhadip@imsc.res.in} }

\date{}

\begin{document}
 \maketitle
\vspace{-0.6cm}
\bc
{\small 1) Department of Physics, IIT Kanpur, \\Kanpur 208 016, India\\
2) The Institute of Mathematical Sciences, \\Chennai 600 113, India}\ec

\centerline{\date{\today}}
\vspace{0.5cm}

\bc
\begin{minipage}{0.8\textwidth}{\small {\bf Abstract:}
We compute the cosmological constant in a scale invariant 
scalar field theory. The gravitational action is also suitably modified
to respect scale invariance. Due to scale invariance the theory 
does not admit a cosmological constant term. The scale invariance
is broken by a recently introduced mechanism called
cosmological symmetry breaking. This leads to a nonzero cosmological
constant. We compute the one loop  corrections to the cosmological 
constant and show that it is finite.
}\end{minipage}
\ec

\vspace{0.5cm}

\section{Introduction}
In a recent series of papers \cite{JM,JMS,AJS}, we have investigated
a scale invariant extension of the standard model. The basic idea
has been introduced earlier by Cheng and collaborators \cite{ChengPRL,ChengKao,Kao}. Phenomenological consequences of this model have also been
studied in Ref. \cite{Huang,Wei}.
The essential idea can be captured by considering a simple model where
we include only one real scalar field besides gravity. Hence here
we consider only this simple model which displays global scale invariance.
The action
for this model may be written as,
\begin{equation}
\mc{S} = \int d^4x \sqrt{-g}\left[\frac12g^{\mu\nu}\partial_\mu\Phi\partial_\nu\Phi
-\frac{\lambda}{4}\Phi^4 -\frac{\beta}{8}\Phi^2R \right],
\end{equation}
where $R$ is the Ricci scalar. The model has no dimensionful parameter. In order to agree with observations
scale invariance has to be broken. In Ref. \cite{JM} we argued that it
is broken by a new phenomenon, which we called cosmological symmetry breaking.
This is inspired by the standard big bang model. Here
the universe is described by a time dependent solution
of the classical equations of motion. At leading order the solution
is just the homogeneous and isotropic Friedmann-Robertson-Walker (FRW) model. We argue that in order
to describe physical phenomenon we need to make a quantum expansion
around this classical background.

In Refs. \cite{JM,JMS} we found that, assuming the FRW metric with
scale factor $a(t)$, this
model has a classical solution,
\begin{equation}
a(t) = a_0 \exp(H_0t)
\label{eq:a(t)}
\end{equation}
with
\begin{equation}
\Phi_{cl} = \eta = \sqrt{{3\beta\over \lambda}} \, H_0 ,
\label{eq:Phi_cl}
\end{equation}
where $\eta$ is the classical solution of the scalar field $\Phi$ and
$H_0$ is the Hubble constant, which is independent of time in the
present case. Here we have set the spatial curvature parameter $k=0$ 
in the FRW metric.

A basic problem with imposing scale invariance is that it is believed to
be anomalous \cite{Coleman,Collins}. When we compute the quantum loop corrections, we necessarily need to
regulate the action, which introduces a scale. Hence one concludes that
scale invariance is broken.
We have argued in Refs. \cite{JM,JMS} that this conclusion need
not hold in the present case. The basic point is that here the classical
solution $\eta$ itself provides a scale. Here we expand the field 
$\Phi$ around this classical solution,
\begin{equation}
\Phi(x) = \eta + \phi(x)\, ,
\end{equation}
where $\phi(x)$ represent the quantum fluctuations. 
Under scale transformations
$\eta$ also scales in exactly the same form as $\Phi$. In dimensional
regularization the regulated action in $d=4-\epsilon$ dimensions
may be written as,
\begin{equation}
\mathcal{S}
= \int d^dx \sqrt{-g}\Bigg({1\over 2}g^{\mu\nu}\partial_\mu\Phi\partial_\nu\Phi
-{\lambda\over 4}\Phi^4 \eta^{-2\delta}
-\frac{\beta}{8}\Phi^2R  \Bigg).
\label{eq:Action_d_dim}
\end{equation}
where $\delta = (d-4)/(d-2)$.
Here we have essentially used the classical field to introduce the
scale required in this action.
It is clear that even after regularization the action is invariant under
scale transformations. Hence we do not expect the scale invariance to be
anomalous. The regularized action, Eq. \ref{eq:Action_d_dim}, is 
slightly different from what we proposed earlier in Ref. \cite{JMS}. A 
detailed comparison between the two will be presented in a 
separate paper. Here we simply point out that the action given in Eq. 
\ref{eq:Action_d_dim} is exactly scale invariant and well defined as
long as $\eta\ne 0$. The details about the transformation rule
are given in Ref. \cite{JMS}. The transformation 
is similar to what was also proposed
in Ref. \cite{Englert}.

An important aspect of a scale invariant theory is that it does not
permit a cosmological constant term in the action. Hence this may potentially
solve the well known cosmological constant problem 
\cite{Weinberg,Peebles,Padmanabhan,Copeland,Carroll,Sahni,Ellis}. 
Alternate approaches to solving the cosmological constant problem are 
described in Refs. \cite{Weinberg,Aurilia,VanDer,Henneaux,Brown,Buchmuller,Henneaux89,Sorkin,Sundrum}.
As shown in Refs. \cite{JM,JMS} a cosmological constant is generated
by the classical solution. Furthermore if the
full quantum theory is truly scale invariant, as we claim, we expect
that the cosmological constant would be finite at all orders in perturbation
theory. Although this might be anticipated \cite{JMS} since the
regulated theory is scale invariant, an explicit demonstration is necessary 
since the regularization procedure \cite{JMS} is somewhat unusual.
In the present paper we demonstrate that at one loop the
cosmological constant is calculable and finite. As far as we know this
is the first demonstration that the loop contributions to the 
cosmological constant may be 
finite in a quantum field theory.

In pure gravity theory with Minkowski background it has earlier been 
found that the one loop divergent contributions to cosmological constant vanish
\cite{Mueller}. However this cancellation does not survive at two
loops \cite{Sawhill}. In the present case, however, we demonstrate
this cancellation including a matter field with the background metric
being the standard FRW metric. Furthermore the cancellation is a result of a
symmetry, i.e. the scale invariance, and hence is expected to extend to all
orders.

In the present paper we shall ignore quantum gravity corrections. These
can be shown to be higher order in powers of $1/\beta$. Hence they can
be ignored consistently.

\section{Renormalization}
In this section we shall only expand the scalar field. For calculation of
the cosmological constant we need to expand the metric also. This will be
done later. Here we simply set the metric equal to the FRW metric with
a scale factor $a(t)$. We shall
assume that the universe is evolving very slowly with time. Hence in the
loop integrals, the time dependent factors $a(t)$ will be set to their current
values. This is essentially an adiabatic approximation.
The Ricci scalar, 
\begin{equation}
R=-12 H_0^2\, .
\label{eq:R}
\end{equation}
Hence, if we ignore quantum gravity
contributions, the term proportional to $R$ simply acts like a mass term.
Therefore we can obtain all the counterterms by simply repeating the standard
field theoretic analysis for a spontaneously broken theory. We caution the
reader that the fundamental mechanism here is very different from
spontaneous symmetry breaking.

The \Lag in $d=4-\ep$ dimension (see  
Eq. \ref{eq:Action_d_dim}) in terms of bare field and parameters may be
expressed as,
\ba
\lag=\frac12 g^{\mu\nu} \pr_\m\Ph'\pr_\n\Ph'  -\frac{\lm_0}{4\et'^{2\delta}}\Ph'^4-\frac{\bt_0R}{8}\Ph'^2,\label{eq:L1}
\ea
where $\Ph'=\et'+\ph'$. Let $\Ph=\et+\ph$ be the renormalized field. This is related to
$\Phi'$ by the field renormalization $Z$,
\ba
\Ph'&=&\sqrt Z\Ph.
\ea
We also have $\eta' = \sqrt{Z} \eta$ and $\phi' = \sqrt{Z}\phi$.
We define the counterterms as,
\ba
\dl_Z&=&Z-1,\label{eq:def_dl_Z}\\
\dl_\lm&=&\lm_0Z^{2(1-\delta)}-\lm,\label{eq:def_dl_l}\\
\dl_\bt&=&\bt_0Z-\bt\label{eq:def_dl_b},
\ea
where $\lm$ and $\bt$ are the physical coupling constants.
We can rearrange the \Lag as,
\ba
\lag =\left[\frac12g^{\mu\nu}\pr_\mu\Ph\pr_\nu\Ph -\frac{\lm}{4\et^{2\delta}}\Ph^4 -
\frac{\bt R}{8}\Ph^2\right]
+\lag_{ct}\, ,
\ea
where the counterterm \Lag is given by,
\ba
\lag_{ct} =
\frac12\dl_Z g^{\mu\nu}\pr_\mu\Ph\pr_\nu\Ph -\frac{\dl_\lm}{4\et^{2\delta}}\Ph^4 -
\frac{\dl_\bt R}{8}\Ph^2.
\ea
We may use the solution of the classical field equation, $\eta$,
to rewrite the \Lag as,
\ba
\nn\lag &=&\left[\frac12g^{\mu\nu}\pr_\m\ph\pr_\nu\ph -\frac{\lm}{4\et^{2\delta}}(\et^4+6\et^2\ph^2+4\et\ph^3+\ph^4) -\frac{\bt R}{8}(\et^2+\ph^2)\right]\\
\nn&+&
\left[\frac12\dl_Z g^{\mu\nu} \pr_\m\ph\pr_\nu\ph -\frac{\dl_\lm}{4\et^{2\delta}}(4\et^3\ph+6\et^2\ph^2+4\et\ph^3+\ph^4) -\frac{\dl_\bt R}{8}(2\et\ph+\ph^2)\right]\\
&-&\left(\frac{\dl_\lm}{4\et^{2\delta}}\et^4+\frac{\dl_\bt R}{8}\et^2\right)\label{eq:L2}.
\ea
Here the terms linear in $\phi$ in the leading order Lagrangian,
i.e. not including counterterms, vanish if 
$\beta R/4= -\lambda\eta^{2-2\delta}$. 
Hence $m_\ph^2=3\lm\et^{2-2\delta}+\bt R/4 = 2\lm\et^{2-2\delta}$ where $m_\ph$ denotes the mass of $\ph.$ We point out that $\eta$ has the same dimensions
as the field $\phi$ and hence its dimension is a little different from the
dimension of mass.
We keep the terms given in the third line of Eq. \ref{eq:L2} as these will be relevant
when we expand the metric.

It is now easy to obtain the Feynman rules for the various vertices and counterterms. In Table \ref{tab:fr} we have listed the Feynman rules.

\begin{table}[!t]
\bc
  \begin{tabular}{rcl}
  \hline\hline
  \multirow{3}{*}{\includegraphics[width=0.10\textwidth]{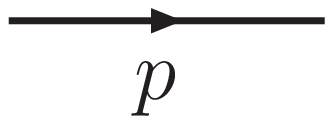}} &\hspace{40mm}&\\
  &\hspace{40mm}&$i/(p^2-m_\ph^2)$\\
  &\hspace{40mm}&\\
  \multirow{3}{*}{\includegraphics[width=0.10\textwidth]{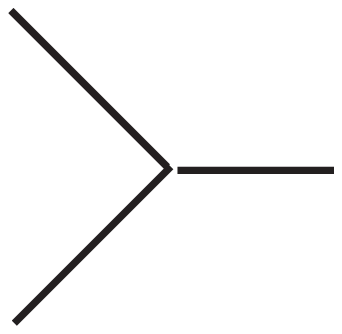}} &\hspace{40mm}&\\
  &\hspace{40mm}&$-6i\lm\et^{1-2\delta}$\\
  &\hspace{40mm}&\\
  \multirow{3}{*}{\includegraphics[width=0.10\textwidth]{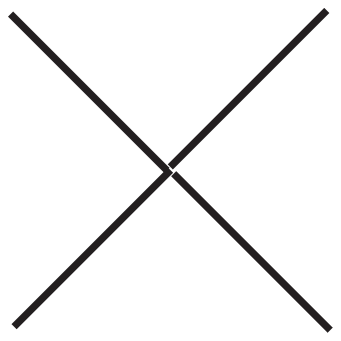}} &\hspace{40mm}&\\
  &\hspace{40mm}&$-6i\lm\et^{-2\delta}$\\
  &\hspace{40mm}&\\
   \multirow{3}{*}{\includegraphics[width=0.10\textwidth]{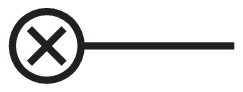}} &\hspace{40mm}&\\
  &\hspace{40mm}&$-i(\dl_\lm\et^{3-2\delta}+\dl_\bt R\et/4)$\\
  &\hspace{40mm}&\\
   \multirow{3}{*}{\includegraphics[width=0.10\textwidth]{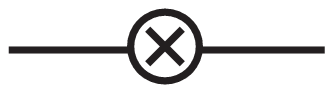}} &\hspace{40mm}&\\
  &\hspace{40mm}&$i(p^2\dl_Z-3\dl_\lm\et^{2-2\delta}-\dl_\bt R/4)$\\
  &\hspace{40mm}&\\
   \multirow{3}{*}{\includegraphics[width=0.10\textwidth]{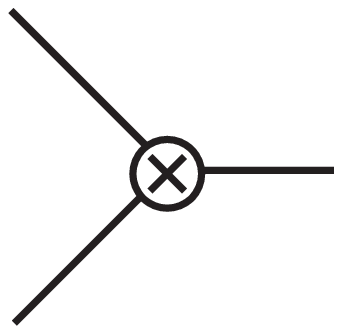}} &\hspace{40mm}&\\
  &\hspace{40mm}&$-6i\dl_\lm\et^{1-2\delta}$\\
  &\hspace{40mm}&\\ \multirow{3}{*}{\includegraphics[width=0.10\textwidth]{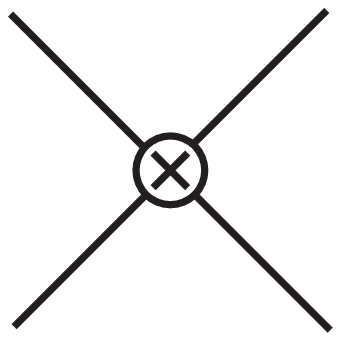}} &\hspace{40mm}&\\
  &\hspace{40mm}&$-6i\dl_\lm\et^{-2\delta}$\\
  &\hspace{40mm}&\\\hline\hline
  \end{tabular}
  \caption{Feynman rules for the scalar field.}\label{tab:fr}
\ec
  \end{table}

\section{One Loop Calculation of Counterterms}

\begin{figure}[!b]\bc
\includegraphics[width=0.4\textwidth]{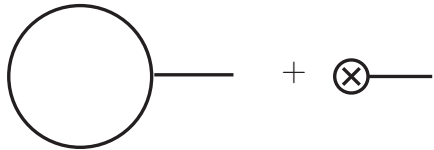}
\caption{1-point contribution at one loop level.}\label{fig:1lp1pt}
\ec\end{figure}

In this section we shall evaluate the counterterms defined in the last section (Eqs. \ref{eq:def_dl_Z}, \ref{eq:def_dl_l}, \ref{eq:def_dl_b})
up to one loop level. We shall impose the condition,
\ba
<\Phi> = \sqrt{{3\beta\over \lambda}} H_0 \eta^\delta
\ea 
at all orders in perturbation theory. Here the symbol $<\Phi>$ means the 
expectation value of the field $\Phi$ in the lowest energy state when
we expand around a nontrivial classical solution. A similar condition
may also be imposed in the case of spontaneous symmetry breaking 
\cite{Peskin}. However in the case of spontaneous breaking
the meaning of $<\Phi>$ is the vacuum expectation
value of $\Phi$, which is different from our case. 
The total 1-point amplitude at one loop level is shown in Fig. \ref{fig:1lp1pt}.
From Table \ref{tab:fr} we get, \ba
i\mc M^{\rm 1pt}_{\rm 1loop} &=& -3i\left(\frac{\lm\et}{\et^{2\delta}}\right)\frac{\G(\ep/2-1)}{(4\pi)^{d/2}}\left(\frac{1}{m_\ph^2}\right)^{\ep/2-1} \label{eq:1lp1pt}.
\ea
We set our renormalization condition such that the total 1-point amplitude vanishes, i.e.,
\bas
-i(\dl_\lm\et^{3-2\delta}+\frac14\dl_\bt R\et)-3i\et\left(\frac{\lm}{\et^{2\delta}}\right)\frac{\G(\ep/2-1)}{(4\pi)^{d/2}}\left(\frac{1}{m_\ph^2}\right)^{\ep/2-1}=0,
\eas
or,
\ba
(\dl_\lm\et^2+\frac14\dl_\bt R \eta^{2\delta})&=&-3\lm\frac{\G(\ep/2-1)}{(4\pi)^{d/2}}\left(\frac{1}{m_\ph^2}\right)^{\ep/2-1}\nn\\
&=&\frac{6\lm^2\et^2}{(4\pi)^2}\left[\frac2\ep+1-\g-\log\left(\frac{\lm}{2\pi}\right)\right].\label{eq:dl_bt}
\ea
The last relation fixes $\dl_\bt$ in terms of $\dl_\lm$.

\begin{figure}[!b]\bc
\includegraphics[width=0.5\textwidth]{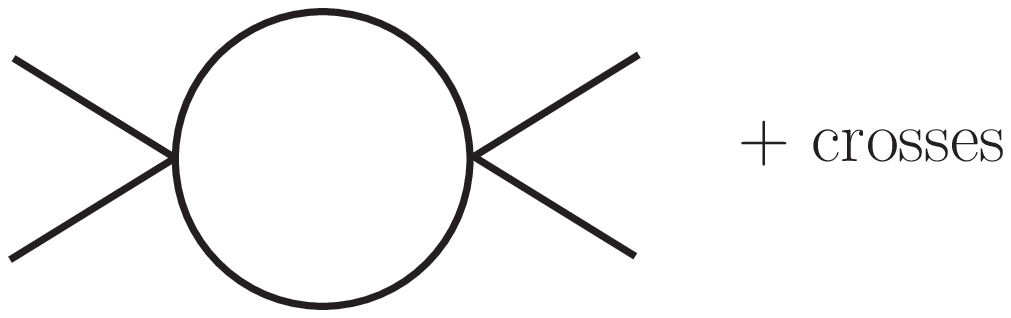}
\caption{4-point divergent contribution at one loop level. }\label{fig:1lp4pt}
\ec\end{figure}

\begin{figure}[!t]\bc
\subfigure[]{\label{fig:1lp4ptlm3}\includegraphics[width=0.45\textwidth]{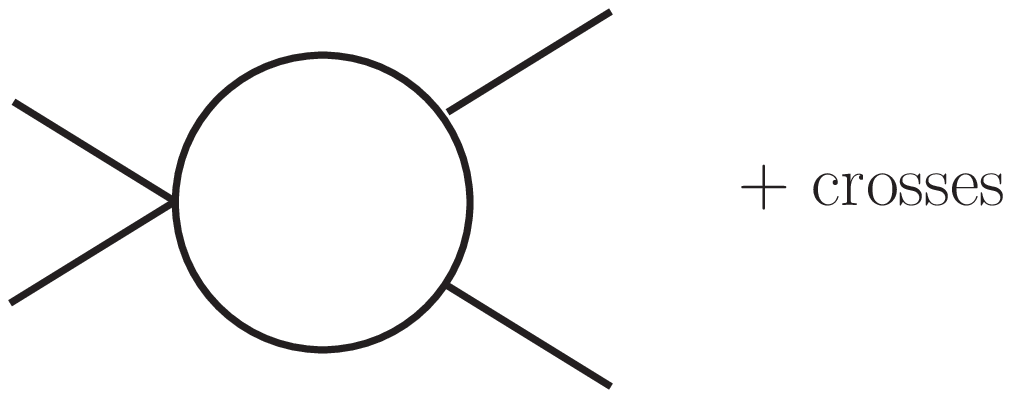}}
\subfigure[]{\label{fig:1lp4ptlm4}\includegraphics[width=0.45\textwidth]{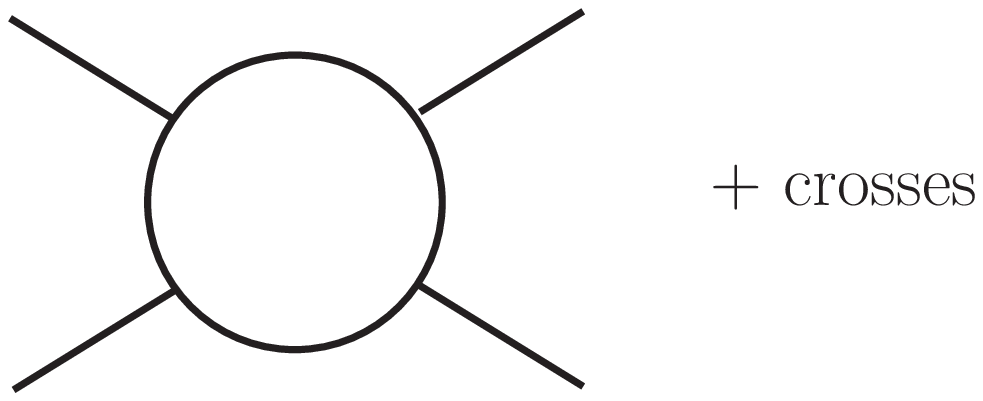}}
\caption{Order $\lm^3$ and $\lm^4$ contributions to the 4-point function.}
\ec\end{figure}

To set $\dl_\lm$ we consider the 4-point function.
At one loop level, the diverging 4-point contributions arise only from the $\mc O(\lm^2)$ diagrams shown in Fig \ref{fig:1lp4pt}.
\ba
i\mc M^{\rm 4pt}_{\rm 1loop} &=& \left(\frac{-6i\lm}{\et^{\delta}}\right)^2\cdot i[I(s)+I(t)+I(u)] +\mc O(\lm^3),\ea
where
\ba
iI(p^2)&=& {1\over 2 \eta^{2\delta}}\int\frac{d^dk}{(2\pi)^d}\frac{i}{k^2-m_\ph^2}\frac{i}{(k+p)^2-m_\ph^2}\nn\\
&=&-\frac i2\frac{\G(\ep/2)}{(4\pi)^2}\int^1_0dz\left[\frac{4\pi\eta^2}{m_\ph^2-z(1-z)p^2}\right]^{\ep/2}\nn\\
&=&-\frac i2\frac{1}{(4\pi)^2}\left[\frac2\ep-\g-\int^1_0 dz\log\left(\frac{m_\ph^2-z(1-z)p^2}{4\pi\eta^2}\right)\right]\\
&\equiv&-\frac i2\frac{1}{(4\pi)^2}\frac2\ep+iI'(p^2).
\ea
The last line defines $I'(p^2)$. Here $\g$ is the Eular-Mascheroni constant. Hence,
\ba
i\mc M^{\rm 4pt}_{\rm 1loop} = 54i\left(\frac{\lm}{\et^{\delta}}\right)^2\frac{1}{(4\pi)^2}\left(\frac2\ep\right)
+\left(-6i\lm\right)^2\cdot i[I'(s)+I'(t)+I'(u)]+\mc O(\lm^3).\label{eq:1lp4pt}
\ea
The order $\lm^3$ and $\lm^4$ contributions come from the diagrams shown in Figs. \ref{fig:1lp4ptlm3} and \ref{fig:1lp4ptlm4}, respectively.
However it is easy to check that they are finite. These are higher order in
$\lambda$ and we shall ignore them. 
We can get rid of the infinite term by demanding
\ba
\dl_\lm =\frac{ 9\lm^2}{(4\pi)^2}\left(\frac2\ep\right) -{6\lm^2}[I'(s_0)+I'(t_0)+I'(u_0)],\label{eq:dl_lm}
\ea
where we have set our renormalization point at $s=s_0$, $t=t_0$ and $u=u_0$.

\begin{figure}[!b]\bc
\includegraphics[width=0.6\textwidth]{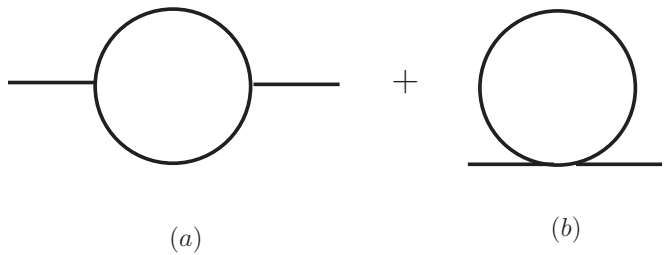}
\caption{One loop correction to the propagator.}\label{fig:1lp2pt}
\ec\end{figure}

Since the 1-point function vanishes at the one loop level we have only two divergent diagrams for the two point amplitude
(Fig. \ref{fig:1lp2pt}). We can write,
\ba
i\mc M^{\rm 2pt}_{\rm 1loop(a)} &=& 18i\left(\frac{\lm\et}{\et^{\delta}}\right)^2\frac{1}{(4\pi)^2}\left(\frac2\ep\right)
+\left(-6i\lm\et\right)^2\cdot iI'(p^2) 
\label{eq:1lp2pt-a},\\
i\mc M^{\rm 2pt}_{\rm 1loop(b)} &=& -3i\left(\frac{\lm}{\et^{2\delta}}\right)\frac{\G(\ep/2-1)}{(4\pi)^{d/2}}\left(\frac{1}{m_\ph^2}\right)^{\ep/2-1} \label{eq:1lp2pt-b}.
\ea
The counterterm for the 2-point amplitude is given by,
\bas
i(p^2\dl_Z-3\dl_\lm\et^{2-2\delta}-\dl_\bt R/4)=ip^2\dl_Z-2i\dl_\lm\et^{2-2\delta}-i(\dl_\lm \et^{2}+\dl_\bt R\eta^{2\delta}/4)\et^{-2\delta}.
\eas
From Eqs. \ref{eq:dl_lm} and \ref{eq:dl_bt} we see that the last two terms in the r.h.s. of the above equation cancel all the singularities. As there
is no divergent term proportional to $p^2$ we set
\ba
\dl_Z=0
\ea
at one-loop level.

\begin{figure}[!t]\bc
\includegraphics[width=0.5\textwidth]{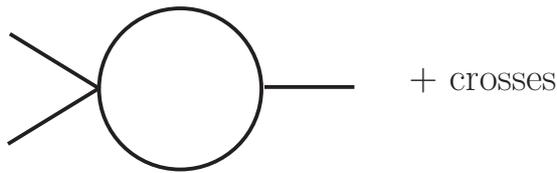}
\caption{3-point divergent contribution at one loop level. }\label{fig:1lp3pt}
\ec\end{figure}
The divergent 3-point contribution comes from the diagrams shown in Fig \ref{fig:1lp3pt}.
It is easy to derive,
\ba
i\mc M^{\rm 3pt}_{\rm 1loop} &=& 54i\et\left(\frac{\lm}{\et^{\delta}}\right)^2\frac{1}{(4\pi)^2}\left(\frac2\ep\right) +\mc F~\label{eq:1lp3pt}.
\ea
where $\mc F$ indicates a finite part.
However this infinity does not give rise
to any problem as the three point counterterm, shown in Table \ref{tab:fr}, cancels it precisely.

Finally we compute the constant terms in the Lagrangian, i.e. the
terms given in the third line of Eq. \ref{eq:L2}. These evaluate to,
\ba
\mc L_{\rm X}&=&-\left(\frac{\dl_\lm}{4\et^{2\delta}}\et^4+\frac{\dl_\bt R}{8}\et^2\right)\nonumber\\ 
&=&-\frac34\left(\frac{\lm\et^2}{4\pi}\right)^2\left(\frac2\ep\right)+\mc F_{\rm X}.
\label{eq:LX}
\ea
where $\mc F_{\rm X}$ denotes the finite part.

\section{Cosmological Constant at One Loop}
\begin{figure}[!t]\bc
\includegraphics[width=0.3\textwidth]{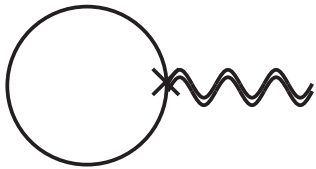}
\caption{One loop correction to cosmological constant. The external
line here represents a graviton.}\label{fig:cos_const_1lp}
\ec\end{figure}

In this section we compute the cosmological constant, $\Lambda$, at one loop.
This is defined by the term in the action
\begin{equation}
S_{\Lambda} = -\int d^dx \sqrt{-\bar{g}} \Lambda
\end{equation}
In our scale invariant theory such a term is absent from the action, 
as discussed above. However due to cosmological breaking of scale invariance
we find a nonzero contribution.
 We expand the metric such that
\begin{equation}
\bar{g}_{\mu\nu} = g_{\mu\nu} + h_{\mu\nu}
\end{equation}
Here we follow the notation of Ref. \cite{tHooft} and denote the full
metric as well as other variables with a {\it bar}. The symbol $g_{\mu\nu}$
represents the classical metric and $h_{\mu\nu}$ the quantum field. In our
case $g_{\mu\nu}$ is simply the FRW metric. Here we shall expand gravity
only to first order in $h_{\mu\nu}$.
 We are only interested in the computation of the cosmological
constant which is identified with the one point function in gravity.
This would involve the computation of the counter lagrangian proportional
to $h=h^\alpha_\alpha$ and the one loop graph
 shown in Fig. \ref{fig:cos_const_1lp}.
At first order in $h$, we find \cite{tHooft},
$\bar{g}^{\mu\nu} = g^{\mu\nu} - h^{\mu\nu}$, $\sqrt{-\bar{g}} =
\sqrt{-{g}}(1+h/2)$ and
\begin{equation}
\bar{R} = R + h_{\beta ;\alpha}^{\beta ;\alpha} -
h_{\alpha ;\ \beta}^{\beta ;\alpha} - h^\nu_\alpha R^\alpha_\nu\, ,
\label{eq:barR}
\end{equation}
where $R_{\mu\nu}$ is the curvature tensor corresponding to the FRW
metric. Here ``$;$" denotes covariant derivatives.

We use the conformal time in the FRW metric. Hence the classical metric becomes,
\ba
g_{\mu\nu} = a^2 \cdot {\rm\ diagonal}(1,-1,-1,-1).
\ea 
The Ricci scalar is given in Eq. \ref{eq:R}
and the tensor,
\begin{equation}
R_{\mu\nu} = -3H_0^2g_{\mu\nu}\, ,
\end{equation}
where $H_0$ is the Hubble constant.

We first consider the counterterm action,
\begin{eqnarray}
S_{ct} &=& \int d^dx \sqrt{-\bar{g}}\left(-{\delta_\beta \bar{R} \over {8}}
{\Phi}^2 -{\delta_\lambda\over {4\et^{2\delta}}} {\Phi}^4\right)\\
&=& \int d^dx \sqrt{-\bar{g}}\left(-{\delta_\beta \bar{R} \over {8}}
{\et}^2 -{\delta_\lambda\over {4\et^{2\delta}}} {\et}^4\right)+\ldots~.
\end{eqnarray}
We identify the terms proportional to $h$ in the
counterterm action. We find, dropping surface terms,
\begin{eqnarray}
S_{ct} &=&\int d^dx \sqrt{-g}\frac h4\left[\lag_{\rm X}-\frac{\dl_\lm\et^4}{4\et^{2\delta}}\right] + \ldots \nonumber\\
&=& -\int d^d x\frac34 \sqrt{-{g}}h \left({\lambda\eta^2}\over {4\pi\eta^\delta}\right)^2\left(\frac2\ep +1-\gamma-\log\left({\lambda\over 2\pi}\right) \right)+\ldots,
\end{eqnarray}
where $\lag_{\rm X}$ is defined in Eq. \ref{eq:LX}. In this equation
we have displayed only the divergent terms.
We next identify the terms proportional to $h\phi^2$ which would
contribute to the one loop diagram with an external $h$ line.
We find
\begin{equation}
- \sqrt{-\bar{g}}{\lambda\over {4\et^{2\delta}}}\Phi^4 = -{3\over 4}\sqrt{-{g}}\lambda\eta^{2-2\delta}
h\phi^2 + \ldots~,
\end{equation}
\begin{equation}
\sqrt{-\bar{g}}{\bar{g}^{\mu\nu}\over 2}\partial_\mu\Phi
\partial_\nu \Phi = {1\over 2}\sqrt{-{g}}\left({h\over 2} g^{\mu\nu}
- h^{\mu\nu}\right)\partial_\mu\phi\partial_\nu \phi
\end{equation}
and
\begin{equation}
- \sqrt{-\bar{g}}{\beta\over {8}}\Phi^2\bar R = -\sqrt{-g}{\bt\over {8}}\frac{R}{4}
h\phi^2 - {\beta\over {8}} \sqrt{-{g}}(\partial_\beta\phi^2)
\left(\Gamma^\alpha_{\alpha\gamma}h^{\gamma\beta} + \Gamma^\beta_{\alpha\gamma}
h^{\alpha\gamma}\right)+\ldots~.
\end{equation}
In the fourier space the second term on the right hand side would be
proportional to $k_\beta$. Here we are working in the adiabatic limit
where the background metric is assumed to be very slowly varying. Hence
the background metric is taken out of the Feynman integral. A Feynman
integral proportional to $k_\beta$ is zero, by symmetric integration. Hence
the second term does not contribute.

We now evaluate the one loop contribution to graviton one point function.
The terms which do not involve derivatives of $\ph$ give the contribution
to the amplitude equal to
\ba
-\sqrt{-{g}}\frac58\left(\frac{\lm\et^2}{\et^{2\delta}}\right)\frac{\G(\ep/2-1)}{(4\pi)^{d/2}}\left(\frac{1}{m^2_\ph}\right)^{\ep/2-1}\hspace{2cm}\nn\\
={5\over 4} \sqrt{-{g}} \left({\lambda\eta^2}\over {4\pi \et^\delta}\right)^2\left[\frac2\ep+1-\g-\log\left(\frac{\lm}{2\pi}\right)\right].
\ea
The KE term for the scalar field gives
\ba
- \sqrt{-{g}} \frac{\G(-d/2)}{4(4\pi)^{d/2}}\left(\frac{1}{m_\ph^2}\right)^{-d/2}
=-{1\over 2} \sqrt{-{g}}\left({\lambda\eta^2}\over {4\pi \et^\delta}\right)^2\left[\frac2\ep+\frac32-\g-\log\left(\frac{\lm}{2\pi}\right)\right].
\ea
These two precisely cancel the divergent contributions coming from the
counterterm action. Hence we find that the cosmological constant
is one loop finite. 
The one loop correction to the cosmological constant is found to be 
\begin{equation}
\delta\Lambda = {1\over 2}\left({\lambda\eta^2\over 4\pi}\right)^2\, .
\end{equation}

\section{Conclusions}
The cosmological constant remains one of the most serious issues in physics.
In quantum field theory it acquires infinite contributions which have
to be cancelled at each order in perturbation theory by adding suitable
counter terms. In this case quantum field theory is unable to predict
its value.
We have earlier hypothesized that scale invariance
might control this parameter. In this case we are not permitted to include
a cosmological constant term in the action. If scale invariance
is unbroken then this parameter would be zero at all orders. We 
argued in earlier papers that 
scale invariance is broken due to cosmological symmetry
breaking \cite{JM,JMS,AJS}. 
This phenomenon generates a nonzero cosmological constant.

We have computed the cosmological constant in a  scale invariant scalar 
field theory. 
Due to scale invariance we expect it to be finite at all orders. We have
explicitly demonstrated this to one loop order in a simple model. 
We expect this to also hold in a scale invariant standard model 
\cite{ChengPRL,ChengKao,Kao}. A particularly interesting extension of
this model is to impose local scale invariance \cite{ChengPRL,ChengKao,Kao,Padmanabhan85,Hochberg,Wood,Wheeler,Feoli,Pawlowski,Nishino,Demir}. This
has the interesting prediction that the standard model Higgs particle
disappears from the particle spectrum. We expect our mechanism to work
in these scale invariant theories also.


\begin{thebibliography}{unsrt}
\bibitem{JM} P.~Jain and S.~Mitra,
  Mod.\ Phys.\ Lett.\  A {\bf 22}, 1651 (2007), hep-ph/0704.2273.

\bibitem{JMS}
  P.~Jain, S.~Mitra and N.~K.~Singh,
  JCAP {\bf 0803}, 011 (2008), hep-ph/0801.2041.

\bibitem{AJS}
  P.~K.~Aluri, P.~Jain and N.~K.~Singh, hep-ph/0810.4421.

\bibitem{ChengPRL}
  H.~Cheng,
  Phys.\ Rev.\ Lett.\  {\bf 61}, 2182 (1988).

\bibitem{ChengKao} H. Cheng and W. F. Kao, MIT preprint Print-88-0907 (1988).

\bibitem{Kao} W. F. Kao, Phys. Lett. {\bf A 154}, 1 (1991).
\bibitem{Huang} C.-g. Huang, D.-d. Wu and H.-q. Zheng,
Commun. Theor. Phys. {\bf 14}, 373 (1990).
\bibitem{Wei} H. Wei and R.-G. Cai, JCAP {\bf 0709}, 015 (2007).
\bibitem{Coleman} S. Coleman and R. Jackiw, Annals of Physics {\bf 67},
552 (1971).
\bibitem{Collins} J. C. Collins, A. Duncan and S. D. Joglekar, 
Phys. Rev. {\bf D16}, 438 (1977).
\bibitem{Englert} F. Englert, C. Truffin and R. Gastmans, Nucl. Phys. 
{\bf B117}, 407 (1976).
\bibitem{Weinberg} S. Weinberg, Rev. Mod. Phys. {\bf 61}, 1 (1989).
\bibitem{Peebles} P. J. E. Peebles and B. Ratra, Rev. Mod. Phys. {\bf 75}, 559
(2003).
\bibitem{Padmanabhan} T. Padmanabhan, Phys. Rep. {\bf 380},
235 (2003).
\bibitem{Copeland} E.~J.~Copeland, M.~Sami and S.~Tsujikawa,
Int. J. Mod. Phys.   {\bf D 15}, 1753 (2006).
\bibitem{Carroll} S. M. Carroll, W. H. Press and E. L. Turner,
Ann. Rev. Astron. Astrophys. {\bf 30}, 499 (1992).
\bibitem{Sahni} V. Sahni and A. A. Starobinsky, Int. J. Mod.
Phys. {\bf D 9}, 373 {2000}.
\bibitem{Ellis} J. R. Ellis, Phil. Trans. Roy. Soc. Lond.
{\bf A 361}, 2607 (2003).
\bibitem{Aurilia} A. Aurilia, H. Nicolai and P.K. Townsend,
Nucl. Phys. {\bf B 176}, 509 (1980).
\bibitem{VanDer} J. J. Van Der Bij,  H. Van Dam
and Y. J. Ng, Physica {\bf A 116}, 307 (1982).
\bibitem{Henneaux} M. Henneaux and C. Teitelboim, Phys. Lett. {\bf B 143},
415 (1984).
\bibitem{Brown} J. D. Brown and C. Teitelboim, Nucl. Phys. {\bf B 297},
787 (1988).
\bibitem{Buchmuller} W. Buchmuller and N. Dragon, Phys. Lett. {\bf B 223},
313 (1989).
\bibitem{Henneaux89} M. Henneaux and C. Teitelboim, Phys. Lett. {\bf B 222}, 195(1989).
\bibitem{Sorkin} A. Daughton, J. Louko and R. D. Sorkin, Talk given at 5th
Canadian Conference on General Relativity and Relativistic Astrophysics
(5CCGRRA), Waterloo, Canada, 13-15 May 1993, published in
Canadian Gen. Rel. 0181, (1993).
\bibitem{Sundrum} D. E. Kaplan and R. Sundrum, JHEP {\bf 0607}, 042 (2006).
\bibitem{Mueller} M. Mueller, Phys. Lett. {\bf B133}, 385 (1983).
\bibitem{Sawhill} B. K. Sawhill, Phys. Lett. {\bf B161}, 112 (1985).
\bibitem{Peskin} {\it An Introduction to Quantum Field Theory}, M. E. Peskin
and D. V. Schroeder, Westview Press (1995).
\bibitem{tHooft}
  G.~'t Hooft and M.~J.~G.~Veltman,
  Annales Poincare Phys.\ Theor.\  A {\bf 20}, 69 (1974).
\bibitem{Padmanabhan85} T. Padmanabhan, Classical and Quantum Gravity,  
{\bf 2}, L105-L108 (1985).
\bibitem{Hochberg} D. Hochberg and G. Plunien, Phys. Rev. {\bf D 43}, 3358 (1991).
\bibitem{Wood} W.R. Wood and  G. Papini, Phys. Rev. {\bf D 5}, 3617 (1992).
\bibitem{Wheeler} J. T. Wheeler, J. Math. Phys. {\bf 39}, 299 (1998).
\bibitem{Feoli} A. Feoli, W.R. Wood and G. Papini,
J. Math. Phys. {\bf 39}, 3322 (1998).
\bibitem{Pawlowski} M. Pawlowski, Turk. J. Phys. {\bf 23}, 895 (1999).
\bibitem{Nishino} H. Nishino and S. Rajpoot, hep-th/0403039.
\bibitem{Demir} D. A. Demir, Phys. Lett. {\bf B 584}, 133 (2004).


\end{thebibliography}
\end{document}